\documentclass[nofootinbib,preprint]{revtex4}%
\usepackage{hyperref}
\usepackage{amsmath}
\usepackage{amsfonts}
\usepackage{amssymb}
\usepackage{graphicx}%
\providecommand{\U}[1]{\protect\rule{.1in}{.1in}}

\begin{document}
\title{The Appell Function $F_{1}$ and Regge String Scattering Amplitudes}
\date{\today}
\author{Jen-Chi Lee}
\email{jcclee@cc.nctu.edu.tw}
\affiliation{Department of Electrophysics, National Chiao-Tung University, Hsinchu, Taiwan, R.O.C.}
\affiliation{Physics Division, National Center for Theoretical Sciences, Hsinchu, Taiwan, R.O.C.}
\author{Yi Yang}
\email{yiyang@mail.nctu.edu.tw}
\affiliation{Department of Electrophysics, National Chiao-Tung University, Hsinchu, Taiwan, R.O.C.}
\affiliation{Physics Division, National Center for Theoretical Sciences, Hsinchu, Taiwan, R.O.C.}

\begin{abstract}
We show that each 26D open bosonic Regge string scattering amplitude (RSSA)
can be expressed in terms of one single Appell function $F_{1}$ in the Regge
limit. This result enables us to derive infinite number of recurrence
relations among RSSA at arbitrary mass levels, which are conjectured to be
related to the known $SL(5,C)$ dynamical symmetry of $F_{1}$. In addition, we
show that these recurrence relations in the Regge limit can be systematically
solved so that all RSSA can be expressed in terms of one amplitude. All these
results are dual to high energy symmetries of fixed angle string scattering
amplitudes discovered previously \cite{ChanLee1,ChanLee2,CHLTY,PRL,susy}.

\end{abstract}
\maketitle

\section{Introduction}

In contrast to the low energy string, the importance of high energy behavior
of string theory was pointed out by Gross \cite{GM, Gross, GrossManes} more
than two decades ago. Recently a saddle point method was invented to
explicitly calculate string scattering amplitudes for string states at
arbitrary mass levels in the fixed angle regime
\cite{ChanLee1,ChanLee2,CHLTY,PRL,susy}. It was found that the ratios of
string scattering amplitudes at each fixed mass level were independent of the
scattering energy and the scattering angle, and the ratios can be extracted at
each mass level . Alternatively, these infinit number of ratios can be
recalculated algebraically by solving linear relations or stringy Ward
identities derived from decoupling of two types of zero-norm states
\cite{ZNS1,ZNS3,ZNS2} in the string spectrum. These infinite linear relations
are so powerful that all fixed angle high energy string scattering amplitudes
can be expressed in terms of one amplitude, say, four tachyon amplitude.

There is another high energy regime of string scattering amplitudes, namely,
the fixed momentum transfer or Regge regime
\cite{RR3,RR4,RR5,RR6,RR7,bosonic,RRsusy}. It was believed that there existed
intimate link between high energy string scattering amplitudes in the fixed
angle regime and in the Regge regime. Note that the number of Regge string
scattering amplitudse (RSSA) is much more numerous than that of high energy
fixed angle string scattering amplitudes. For example, there are only $4$ high
energy fixed angle string scattering amplitudes while there are $22$ RSSA at
mass level $M^{2}=4$ \cite{bosonic}. More recently \cite{LY}, it was
discovered that each RSSA can be expressed in terms of a finite sum of Kummer
functions. One can then solve these Kummer functions at each mass level and
express them in terms of RSSA. Recurrence relations of Kummer functions can
then be used to derive some recurrence relations among RSSA \cite{LY}.
Recurrence relations of higher spin generalization of the BPST vertex
operators \cite{RR6} can also be constructed \cite{Tan}.

Since in general each RSSA was expressed in terms of more than one Kummer
function, it was awkward to derive the complete recurrence relations at
arbitrary higher mass levels. In this letter, we show that each 26D open
bosonic RSSA can be expressed in terms of one single Appell function $F_{1}$.
In contrast to the case of a sum of Kummer functions, this result enables us
to derive the complete infinite number of recurrence relations among RSSA at
arbitrary mass levels, which are conjectured to be related to the known
$SL(5,C)$ dynamical symmetry of $F_{1}$ \cite{sl5c}. In addition, we show that
these recurrence relations in the Regge limit can be systematically solved so
that all RSSA can be expressed in terms of one amplitude. All these results
are dual to high energy symmetries of fixed angle string scattering amplitudes
discovered previously \cite{ChanLee1,ChanLee2,CHLTY,PRL,susy}.

\section{Regge String Scattering Amplitudes}

We first review recent results for high energy string scatterings in the fixed
angle regime,%
\begin{equation}
s,-t\rightarrow\infty,t/s\approx-\sin^{2}\frac{\phi}{2}=\text{fixed (but }%
\phi\neq0\text{)}%
\end{equation}
where $s=-(k_{1}+k_{2})^{2},$ $t=-(k_{2}+k_{3})^{2}$ and $u=-(k_{1}+k_{3}%
)^{2}$ are the Mandelstam variables and $\phi$ is the center of mass
scattering angle. It was shown \cite{PRL,CHLTY} that, at mass level $M_{2}%
^{2}=2(N-1)$, states to the leading order in energy are of the form (the
second state of the four point function is chosen to be the higher spin string
state)
\begin{equation}
\left\vert N,2m,q\right\rangle \equiv(\alpha_{-1}^{T})^{N-2m-2q}(\alpha
_{-1}^{L})^{2m}(\alpha_{-2}^{L})^{q}|0,k\rangle\label{number}%
\end{equation}
where the polarizations of the 2nd particle with momentum $k_{2}$ on the
scattering plane were defined to be $e^{P}=\frac{1}{M_{2}}(E_{2}%
,\mathrm{k}_{2},0)=\frac{k_{2}}{M_{2}}$, $e^{L}=\frac{1}{M_{2}}(\mathrm{k}%
_{2},E_{2},0)$ and $e^{T}=(0,0,1)$. $\eta_{\mu\nu}=diag(-1,1,1)$. $N,m$ and
$q$ in Eq.(\ref{number}) are non-negative integers and $N\geq2m+2q$. Since
$e^{P}$ approaches to $e^{L}$ in the fixed angle regime \cite{ChanLee2}, we
did not put $e^{P}$ components in Eq.(\ref{number}). For simplicity, we choose
the particles associated with momenta $k_{1}$, $k_{3}$ and $k_{4}$ to be
tachyons. The $s-t$ channel high energy fixed angle string scattering
amplitudes can be calculated \cite{PRL} to be
\begin{equation}
\mathcal{T}^{(N,2m,q)}=\sqrt{\frac{2\pi}{Kf_{0}^{\prime\prime}}}e^{-Kf_{0}%
}\left[  (-1)^{N-q}\frac{2^{N-q-2m}(2m)!}{m!\ {M}_{2}^{q+2m}}\ \tau^{-\frac
{N}{2}}(1-\tau)^{\frac{3N}{2}}E^{N}+O(E^{N-2})\right]  \label{GRAM}%
\end{equation}
where $K\equiv-k_{1}.k_{2}\rightarrow2E^{2},$ $f(x)\equiv\ln x-\tau\ln(1-x)$
$,\tau\equiv-\frac{k_{2}.k_{3}}{k_{1}.k_{2}}\rightarrow\sin^{2}\frac{\phi}{2}%
$, and the saddle point for the integration of moduli is defined by
$f^{\prime}(x_{0})=0$ with $x_{0}=\frac{1}{1-\tau}$. The ratios among high
energy fixed angle string scattering amplitudes of different string states at
each fixed mass level can be extracted from Eq.(\ref{GRAM}) to be \cite{PRL}%
\begin{equation}
\frac{T^{(N,2m,q)}}{T^{(N,0,0)}}=\left(  -\frac{1}{M_{2}}\right)
^{2m+q}\left(  \frac{1}{2}\right)  ^{m+q}(2m-1)!!.
\end{equation}
Alternatively, it was discovered that \cite{PRL,CHLTY} the ratios above can be
recalculated by using the decoupling of two types of high energy fixed angle
zero norm states
\begin{align}
L_{-1}\left\vert n-1,2m-1,q\right\rangle  &  \simeq M\left\vert
n,2m,q\right\rangle +(2m-1)\left\vert n,2m-2,q+1\right\rangle ,\label{HZNS1}\\
L_{-2}\left\vert n-2,0,q\right\rangle  &  \simeq\frac{1}{2}\left\vert
n,0,q\right\rangle +M\left\vert n,0,q+1\right\rangle . \label{HZNS2}%
\end{align}
Eqs.(\ref{HZNS1}) and (\ref{HZNS2}) give infinite number of linear relations
among high energy fixed angle string scattering amplitudes of different string
states at each fixed mass level. It turned out that these linear relations can
be systematically solved so that all high energy fixed angle string scattering
amplitudes can be expressed in terms of four tachyon amplitude
\cite{ChanLee1,ChanLee2,CHLTY,PRL,susy}.

We now turn to another high energy regime of string scatterings, namely the
Regge regime, which contains complemantary information \cite{bosonic} of the
theory. That is in the kinematic regime%
\begin{equation}
s\rightarrow\infty,-t=\text{fixed (but }-t\neq\infty).
\end{equation}
It was found \cite{bosonic} that the number of high energy scattering
amplitudes for each fixed mass level in the Regge regime is much more numerous
than that of fixed angle regime in Eq.(\ref{number}) . The leading order high
energy open string states in the Regge regime at each fixed mass level
$N=\sum_{n,m,l>0}np_{n}+mq_{m}+lr_{l}$ are \cite{LY}%
\begin{equation}
\left\vert p_{n},q_{m},r_{l}\right\rangle =\prod_{n>0}(\alpha_{-n}^{T}%
)^{p_{n}}\prod_{m>0}(\alpha_{-m}^{P})^{q_{m}}\prod_{l>0}(\alpha_{-l}%
^{L})^{r_{l}}|0,k\rangle.
\end{equation}
The momenta of the four particles on the scattering plane are%

\begin{align}
k_{1}  &  =\left(  +\sqrt{p^{2}+M_{1}^{2}},-p,0\right)  ,\\
k_{2}  &  =\left(  +\sqrt{p^{2}+M_{2}^{2}},+p,0\right)  ,\\
k_{3}  &  =\left(  -\sqrt{q^{2}+M_{3}^{2}},-q\cos\phi,-q\sin\phi\right)  ,\\
k_{4}  &  =\left(  -\sqrt{q^{2}+M_{4}^{2}},+q\cos\phi,+q\sin\phi\right)
\end{align}
where $p\equiv\left\vert \mathrm{\vec{p}}\right\vert $, $q\equiv\left\vert
\mathrm{\vec{q}}\right\vert $ and $k_{i}^{2}=-M_{i}^{2}$. The relevant
kinematics in the Regge regime are%
\begin{equation}
e^{P}\cdot k_{1}\simeq-\frac{s}{2M_{2}},\text{ \ }e^{P}\cdot k_{3}\simeq
-\frac{\tilde{t}}{2M_{2}}=-\frac{t-M_{2}^{2}-M_{3}^{2}}{2M_{2}};
\end{equation}%
\begin{align}
e^{L}\cdot k_{1}  &  \simeq-\frac{s}{2M_{2}},\text{ \ }e^{L}\cdot k_{3}%
\simeq-\frac{\tilde{t}^{\prime}}{2M_{2}}=-\frac{t+M_{2}^{2}-M_{3}^{2}}{2M_{2}%
};\\
e^{T}\cdot k_{1}  &  =0\text{, \ \ }e^{T}\cdot k_{3}\simeq-\sqrt{-{t}}%
\end{align}
where $\tilde{t}$ $=t-M_{2}^{2}-M_{3}^{2}$ and $\tilde{t}^{\prime}=t+M_{2}%
^{2}-M_{3}^{2}$ . The $s-t$ channel one higher spin and three tachyons string
scattering amplitudes in the Regge limit can be calculated as%
\begin{align}
A^{(p_{n};q_{m};r_{l})}  &  =\int_{0}^{1}dx\,x^{k_{1}\cdot k_{2}}%
(1-x)^{k_{2}\cdot k_{3}}\cdot\left[  \frac{e^{P}\cdot k_{1}}{x}-\frac
{e^{P}\cdot k_{3}}{1-x}\right]  ^{q_{1}}\left[  \frac{e^{L}\cdot k_{1}}%
{x}+\frac{e^{L}\cdot k_{3}}{1-x}\right]  ^{r_{1}}\nonumber\\
&  \cdot\prod_{n=1}\left[  \frac{(n-1)!e^{T}\cdot k_{3}}{(1-x)^{n}}\right]
^{p_{n}}\prod_{m=2}\left[  \frac{(m-1)!e^{P}\cdot k_{3}}{(1-x)^{m}}\right]
^{q_{m}}\prod_{l=2}\left[  \frac{(l-1)!e^{L}\cdot k_{3}}{(1-x)^{l}}\right]
^{r_{l}}\nonumber\\
&  =\prod_{n=1}\left[  (n-1)!\sqrt{-t}\right]  ^{p_{n}}\prod_{m=1}\left[
-(m-1)!\dfrac{\tilde{t}}{2M_{2}}\right]  ^{q_{m}}\prod_{l=1}\left[
(l-1)!\dfrac{\tilde{t}^{\prime}}{2M_{2}}\right]  ^{r_{l}}\nonumber\\
&  \cdot\sum_{j=0}^{r_{1}}\sum_{i=0}^{q_{1}}\binom{r_{1}}{j}{\binom{q_{1}}{i}%
}\left(  -\dfrac{s}{\tilde{t}}\right)  ^{i}\left(  -\dfrac{s}{\tilde
{t}^{\prime}}\right)  ^{j}B\left(  -\frac{s}{2}+N-1-i-j,-\frac{t}%
{2}-1+i+j\right)
\end{align}
where in the Regge limit the beta function $B$ can be further reduced to%
\begin{align}
&  B\left(  -\frac{s}{2}-1+N-i-j,-\frac{t}{2}-1+i+j\right) \nonumber\\
&  \simeq B\left(  -\frac{s}{2}-1,-\frac{t}{2}-1\right)  \dfrac{\left(
-1\right)  ^{i+j}\left(  -\frac{t}{2}-1\right)  _{i+j}}{\left(  \frac{s}%
{2}\right)  _{i+j}}.
\end{align}
Thus%
\begin{align}
A^{(p_{n};q_{m};r_{l})}  &  =B\left(  -\frac{s}{2}-1,-\frac{t}{2}-1\right)
\nonumber\\
&  \cdot\prod_{n=1}\left[  (n-1)!\sqrt{-t}\right]  ^{p_{n}}\prod_{m=1}\left[
-(m-1)!\dfrac{\tilde{t}}{2M_{2}}\right]  ^{q_{m}}\prod_{l=1}\left[
(l-1)!\dfrac{\tilde{t}^{\prime}}{2M_{2}}\right]  ^{r_{l}}\nonumber\\
&  \cdot\sum_{j=0}^{r_{1}}\sum_{i=0}^{q_{1}}\binom{r_{1}}{j}{\binom{q_{1}}{i}%
}\dfrac{\left(  -\frac{t}{2}-1\right)  _{i+j}}{\left(  \frac{s}{2}\right)
_{i+j}}\left(  \dfrac{s}{\tilde{t}}\right)  ^{i}\left(  \dfrac{s}{\tilde
{t}^{\prime}}\right)  ^{j}%
\end{align}
in which the double summation can be expressed in terms of the Appell function
$F_{1}$ as%
\begin{align}
&  \sum_{j=0}^{r_{1}}\sum_{i=0}^{q_{1}}\binom{r_{1}}{j}{\binom{q_{1}}{i}%
}\dfrac{\left(  -\frac{t}{2}-1\right)  _{i+j}}{\left(  \frac{s}{2}\right)
_{i+j}}\left(  \dfrac{s}{\tilde{t}}\right)  ^{i}\left(  \dfrac{s}{\tilde
{t}^{\prime}}\right)  ^{j}\nonumber\\
&  =\sum_{j=0}^{r_{1}}\sum_{i=0}^{q_{1}}\dfrac{\left(  -q_{1}\right)
_{i}\left(  -r_{1}\right)  _{j}}{i!j!}\dfrac{\left(  -\frac{t}{2}-1\right)
_{i+j}}{\left(  \frac{s}{2}\right)  _{i+j}}\left(  -\dfrac{s}{\tilde{t}%
}\right)  ^{i}\left(  -\dfrac{s}{\tilde{t}^{\prime}}\right)  ^{j}\nonumber\\
&  =F_{1}\left(  -\frac{t}{2}-1;-q_{1},-r_{1};\frac{s}{2};-\dfrac{s}{\tilde
{t}},-\dfrac{s}{\tilde{t}^{\prime}}\right)  . \label{finite}%
\end{align}
The Appell function $F_{1}$ is one of the four extensions of the
hypergeometric function $_{2}F_{1}$to two variables and is defined to be%
\begin{equation}
F_{1}\left(  a;b,b^{\prime};c;x,y\right)  =\sum_{m=0}^{\infty}\sum
_{n=0}^{\infty}\dfrac{\left(  a\right)  _{m+n}\left(  b\right)  _{m}\left(
b^{\prime}\right)  _{n}}{m!n!\left(  c\right)  _{m+n}}x^{m}y^{n}%
\end{equation}
where $(a)_{n}=a\cdot\left(  a+1\right)  \cdots\left(  a+n-1\right)  $ is the
rising Pochhammer symbol. Note that when $a$ or $b(b^{\prime})$ is a
nonpositive integer, the Appell function truncates to a polynomial. This is
the case for the Appell function in the RSSA calculated in Eq.(\ref{main}) in
the following%
\begin{align}
A^{(p_{n};q_{m};r_{l})}  &  =\prod_{n=1}\left[  (n-1)!\sqrt{-t}\right]
^{p_{n}}\prod_{m=1}\left[  -(m-1)!\dfrac{\tilde{t}}{2M_{2}}\right]  ^{q_{m}%
}\prod_{l=1}\left[  (l-1)!\dfrac{\tilde{t}^{\prime}}{2M_{2}}\right]  ^{r_{l}%
}\nonumber\\
&  \cdot F_{1}\left(  -\frac{t}{2}-1;-q_{1},-r_{1};\frac{s}{2};-\dfrac
{s}{\tilde{t}},-\dfrac{s}{\tilde{t}^{\prime}}\right)  \cdot B\left(  -\frac
{s}{2}-1,-\frac{t}{2}-1\right)  . \label{main}%
\end{align}
Alternatively, it is interesting to note that the result calculated in
Eq.(\ref{main}) can be directly\ obtained from an integral representation of
$F_{1}$ due to Emile Picard (1881) \cite{Picard}%
\begin{equation}
F_{1}\left(  a;b_{1},b_{2};c;x,y\right)  =\frac{\Gamma(c)}{\Gamma
(a)\Gamma(c-a)}\int_{0}^{1}dt\,t^{a-1}(1-t)^{c-a-1}(1-xt)^{-b_{1}%
}(1-yt)^{-b_{2}}, \label{picard}%
\end{equation}
which was later generalized by Appell and Kampe de Feriet (1926) \cite{Appell}
to $n$ variables%
\begin{align}
F_{1}\left(  a;b_{1},b_{2}...,b_{n};c;x_{1},x_{2}...,x_{n}\right)   &
=\frac{\Gamma(c)}{\Gamma(a)\Gamma(c-a)}\int_{0}^{1}dt\,t^{a-1}(1-t)^{c-a-1}%
\nonumber\\
&  \cdot(1-x_{1}t)^{-b_{1}}(1-x_{2}t)^{-b_{2}}...(1-x_{n}t)^{-b_{n}}.
\label{Kam}%
\end{align}
Eq.(\ref{Kam}) may have application for higher point RSSA \cite{LYi}. To apply
the Picard formula in Eq.(\ref{picard}), we do the transformation
$x\rightarrow\left(  1-x\right)  $, and RSSA can be calculated to be%
\begin{align}
A^{(p_{n};q_{m};r_{l})}  &  =\int_{0}^{1}dx\,\left(  1-x\right)  ^{-\tfrac
{s}{2}+N-2}x^{-\tfrac{t}{2}-2}\cdot\left[  1-\frac{s}{\tilde{t}}\dfrac{x}%
{1-x}\right]  ^{q_{1}}\left[  1-\frac{s}{\tilde{t}^{\prime}}\dfrac{x}%
{1-x}\right]  ^{r_{1}}\nonumber\\
&  \cdot\prod_{n=1}\left[  (n-1)!\sqrt{-t}\right]  ^{p_{n}}\prod_{m=1}\left[
-(m-1)!\dfrac{\tilde{t}}{2M_{2}}\right]  ^{q_{m}}\prod_{l=1}\left[
(l-1)!\dfrac{\tilde{t}^{\prime}}{2M_{2}}\right]  ^{r_{l}}\nonumber\\
&  \simeq B\left(  -\tfrac{t}{2}-1,-\frac{s}{2}-1\right)  \cdot F_{1}\left(
-\tfrac{t}{2}-1,-q_{1},-r_{1},-\tfrac{s}{2};\dfrac{s}{\tilde{t}},\dfrac
{s}{\tilde{t}^{\prime}}\right) \nonumber\\
&  \cdot\prod_{n=1}\left[  (n-1)!\sqrt{-t}\right]  ^{p_{n}}\prod_{m=1}\left[
-(m-1)!\dfrac{\tilde{t}}{2M_{2}}\right]  ^{q_{m}}\prod_{l=1}\left[
(l-1)!\dfrac{\tilde{t}^{\prime}}{2M_{2}}\right]  ^{r_{l}},
\end{align}
which is consistent with the result calculated in Eq.(\ref{main}). It is
important to note that although $F_{1}$ in Eq.(\ref{main}) is a polynomial in
$s$, the result in Eq.(\ref{main}) is valid only for the \textit{leading
order} in $s$ \ in the Regge limit. Note that in contrast to the previous
calculation \cite{LY} in Eq.(\ref{LY}) where a finite sum of Kummer functions
was obtained, here we get only one single Appell function in Eq.(\ref{main}).
This simplification will greatly simplify the calculation of recurrence
relations among RSSA to be discussed in the next section.

\section{Recurrence Relations}

The Appell function $F_{1}$ entails four recurrence relations among contiguous
functions%
\begin{align}
\left(  a-b_{1}-b_{2}\right)  F_{1}\left(  a;b_{1},b_{2};c;x,y\right)
-aF_{1}\left(  a+1;b_{1},b_{2};c;x,y\right)   & \nonumber\\
+b_{1}F_{1}\left(  a;b_{1}+1,b_{2};c;x,y\right)  +b_{2}F_{1}\left(
a;b_{1},b_{2}+1;c;x,y\right)   &  =0,\label{Re1}\\
cF_{1}\left(  a;b_{1},b_{2};c;x,y\right)  -\left(  c-a\right)  F_{1}\left(
a;b_{1},b_{2};c+1;x,y\right)   & \nonumber\\
-aF_{1}\left(  a+1;b_{1},b_{2};c+1;x,y\right)   &  =0,\label{Re2}\\
cF_{1}\left(  a;b_{1},b_{2};c;x,y\right)  +c\left(  x-1\right)  F_{1}\left(
a;b_{1}+1,b_{2};c;x,y\right)   & \nonumber\\
-\left(  c-a\right)  xF_{1}\left(  a;b_{1}+1,b_{2};c+1;x,y\right)   &
=0,\label{Re3}\\
cF_{1}\left(  a;b_{1},b_{2};c;x,y\right)  +c\left(  y-1\right)  F_{1}\left(
a;b_{1},b_{2}+1;c;x,y\right)   & \nonumber\\
-\left(  c-a\right)  yF_{1}\left(  a;b_{1},b_{2}+1;c+1;x,y\right)   &  =0.
\label{Re4}%
\end{align}
All other recurrence relations can be deduced from these four relations. We
can easily solve the Appell function in Eq.(\ref{main}) and express it in
terms of the RSSA%
\begin{align}
&  F_{1}\left(  -\frac{t}{2}-1;-q_{1},-r_{1};\frac{s}{2};-\dfrac{s}{\tilde{t}%
},-\dfrac{s}{\tilde{t}^{\prime}}\right) \nonumber\\
&  =\dfrac{A^{(p_{n};q_{m};r_{l})}}{B\left(  -\frac{s}{2}-1,-\frac{t}%
{2}-1\right)  }\prod_{n=1}\left[  (n-1)!\sqrt{-t}\right]  ^{-p_{n}}\prod
_{m=1}\left[  -(m-1)!\dfrac{\tilde{t}}{2M_{2}}\right]  ^{-q_{m}}\prod
_{l=1}\left[  (l-1)!\dfrac{\tilde{t}^{\prime}}{2M_{2}}\right]  ^{-r_{l}}.
\label{F1}%
\end{align}
Note that among the set of integers $(p_{n},q_{m},r_{l})$ on the right hand
side of Eq.(\ref{F1}), only $(-q_{1},-r_{1})$ dependence shows up on the
Appell function $F_{1}$ on the left hand side of Eq.(\ref{F1}). Indeed, for
those highest spin string states at the mass level $M_{2}^{2}=2\left(
N-1\right)  $%
\begin{equation}
\left\vert N;q_{1},r_{1}\right\rangle \equiv\left(  \alpha_{-1}^{T}\right)
^{N-q_{1}-r_{1}}\left(  \alpha_{-1}^{P}\right)  ^{q_{1}}\left(  \alpha
_{-1}^{L}\right)  ^{r_{1}}|0,k\rangle,
\end{equation}
the string amplitudes reduce to%
\begin{align}
A^{(N;q_{1},r_{1})}  &  =\left(  \sqrt{-t}\right)  ^{N-q_{1}-r_{1}}\left(
-\dfrac{\tilde{t}}{2M_{2}}\right)  ^{q_{1}}\left(  \dfrac{\tilde{t}^{\prime}%
}{2M_{2}}\right)  ^{r_{1}}\nonumber\\
&  \cdot F_{1}\left(  -\frac{t}{2}-1;-q_{1},-r_{1};\frac{s}{2};-\dfrac
{s}{\tilde{t}},-\dfrac{s}{\tilde{t}^{\prime}}\right)  B\left(  -\frac{s}%
{2}-1,-\frac{t}{2}-1\right)  ,
\end{align}
which can be used to solve easily the Appell functin $F_{1}$ in terms of the
RSSA $A^{(N;q_{1},r_{1})}.$

We now proceed to show that the recurrence relations of the Appell function
$F_{1}$ \textit{in the Regge limit} in\textit{ }Eq.(\ref{main}) can be
systematically solved so that all RSSA can be expressed in terms of one
amplitude. As the first step, we note that in \cite{LY} the RSSA was expressed
in terms of finite sum of Kummer functions. There are two equivalent
expressions \cite{LY}%
\begin{align}
A^{(p_{n};q_{m};r_{l})}  &  =\prod_{n>0}\left[  \left(  n-1\right)  !\sqrt
{-t}\right]  ^{p_{n}}\cdot\prod_{m>0}\left[  -\left(  m-1\right)
!\frac{\tilde{t}}{2M_{2}}\right]  ^{q_{m}}\cdot\prod_{l>1}\left[  \left(
l-1\right)  !\frac{\tilde{t}^{\prime}}{2M_{2}}\right]  ^{r_{l}}\nonumber\\
&  \quad\cdot B\left(  -\frac{s}{2}-1,-\frac{t}{2}+1\right)  \left(  \frac
{1}{M_{2}}\right)  ^{r_{1}}\nonumber\\
&  \cdot\sum_{i=0}^{q_{1}}\binom{q_{1}}{i}\left(  \frac{2}{\tilde{t}}\right)
^{i}\left(  -\frac{t}{2}-1\right)  _{i}U\left(  -r_{1},\frac{t}{2}%
+2-i-r_{1},\frac{\tilde{t}^{\prime}}{2}\right) \label{LY2}\\
&  =\prod_{n>0}\left[  \left(  n-1\right)  !\sqrt{-t}\right]  ^{p_{n}}%
\cdot\prod_{m>1}\left[  -\left(  m-1\right)  !\frac{\tilde{t}}{2M}\right]
^{q_{m}}\cdot\prod_{l>0}\left[  \left(  l-1\right)  !\frac{\tilde{t}^{\prime}%
}{2M}\right]  ^{r_{l}}\nonumber\\
&  \cdot B\left(  -\frac{s}{2}-1,-\frac{t}{2}+1\right)  \left(  -\frac
{1}{M_{2}}\right)  ^{q_{1}}\nonumber\\
&  \cdot\sum_{j=0}^{r_{1}}\binom{r_{1}}{j}\left(  \frac{2}{\tilde{t}^{\prime}%
}\right)  ^{j}\left(  -\frac{t}{2}-1\right)  _{j}U\left(  -q_{1},\frac{t}%
{2}+2-j-q_{1},\frac{\tilde{t}}{2}\right)  . \label{LY}%
\end{align}
It is easy to see that, for $q_{1}=0$ or $r_{1}=0$, the RSSA can be expressed
in terms of only one single Kummer function $U\left(  -r_{1},\frac{t}%
{2}+2-i-r_{1},\frac{\tilde{t}^{\prime}}{2}\right)  $ or $U\left(  -q_{1}%
,\frac{t}{2}+2-j-q_{1},\frac{\tilde{t}}{2}\right)  $, which are thus related
to the Appell function $F_{1}\left(  -\frac{t}{2}-1;0,-r_{1};\frac{s}%
{2};-\dfrac{s}{\tilde{t}},-\dfrac{s}{\tilde{t}^{\prime}}\right)  $ or
$F_{1}\left(  -\frac{t}{2}-1;-q_{1},0;\frac{s}{2};-\dfrac{s}{\tilde{t}%
},-\dfrac{s}{\tilde{t}^{\prime}}\right)  $ respectively in the Regge limit
in\textit{ }Eq.(\ref{main}). Indeed, one can easily calculate%
\begin{align}
\lim_{s\rightarrow\infty}F_{1}\left(  -\frac{t}{2}-1;0,-r_{1};\frac{s}%
{2};-\dfrac{s}{\tilde{t}},-\dfrac{s}{\tilde{t}^{\prime}}\right)   &  =\left(
\frac{2}{\tilde{t}^{\prime}}\right)  ^{r_{1}}U\left(  -r_{1},\frac{t}%
{2}+2-r_{1},\frac{\tilde{t}^{\prime}}{2}\right)  ,\\
\lim_{s\rightarrow\infty}F_{1}\left(  -\frac{t}{2}-1;-q_{1},0;\frac{s}%
{2};-\dfrac{s}{\tilde{t}},-\dfrac{s}{\tilde{t}^{\prime}}\right)   &  =\left(
\frac{2}{\tilde{t}}\right)  ^{q_{1}}U\left(  -q_{1},\frac{t}{2}+2-q_{1}%
,\frac{\tilde{t}}{2}\right)  .
\end{align}
On the other hand, it was shown in \cite{LY} that the ratio%
\begin{equation}
\frac{U(\alpha,\gamma,z)}{U(0,z,z)}=f(\alpha,\gamma,z),\alpha=0,-1,-2,-3,...
\label{Lemma}%
\end{equation}
is determined and $f(\alpha,\gamma,z)$ can be calculated by using recurrence
relations of $U(\alpha,\gamma,z)$. Note that $U(0,z,z)=1$ by explicit
calculation. We thus conclude that in the Regge limit%
\begin{equation}
c=\dfrac{s}{2}\rightarrow\infty;x,y\rightarrow\infty;a,b_{1},b_{2}\text{
fixed,}%
\end{equation}
the Appell functions $F_{1}\left(  a;0,b_{2};c;x,y\right)  $ and $F_{1}\left(
a;b_{1},0;c;x,y\right)  $ are determined up to an overall factor by recurrence
relations. The next step is to derive the recurrence relation%
\begin{equation}
yF_{1}\left(  a;b_{1},b_{2};c;x,y\right)  -xF_{1}\left(  a;b_{1}%
+1,b_{2}-1;c;x,y\right)  +\left(  x-y\right)  F_{1}\left(  a;b_{1}%
+1,b_{2};c;x,y\right)  =0, \label{1re}%
\end{equation}
which can be obtained from Eq.(\ref{Re3}) and Eq.(\ref{Re4}). We are now ready
to show that the recurrence relations of the Appell function $F_{1}$
\textit{in the Regge limit} in\textit{ }Eq.(\ref{main}) can be systematically
solved so that all RSSA can be expressed in terms of one amplitude. We will
use the short notation $F_{1}\left(  a;b_{1},b_{2};c;x,y\right)  =F_{1}\left(
b_{1},b_{2}\right)  $ in the following. For $b_{2}=-1$, by using
Eq.(\ref{1re}) and the known $F_{1}\left(  b_{1},0\right)  $ and $F_{1}\left(
0,b_{2}\right)  $, one can easily show that $F_{1}\left(  b_{1},-1\right)  $
are determined for all $b_{1}=-1,-2,-3...$. Similarly, $F_{1}\left(
b_{1},-2\right)  $ are determined for all $b_{1}=-1,-2,-3...$.if one uses the
result of $F_{1}\left(  b_{1},-1\right)  $ in addition to Eq.(\ref{1re}) and
the known $F_{1}\left(  b_{1},0\right)  $ and $F_{1}\left(  0,b_{2}\right)  $.
This process can be continued and one ends up with the result that
$F_{1}\left(  b_{1},b_{2}\right)  $ are determined for all $b_{1}%
,b_{2}=-1,-2,-3...$. This completes the proof that the recurrence relations of
the Appell function $F_{1}$ in the Regge limit in Eq.(\ref{main}) can be
systematically solved so that all RSSA can be expressed in terms of one amplitude.

With the result calculated in Eq.(\ref{main}), one can easily derive many
recurrence relations among RSSA at arbitrary mass levels. For example, the
identity in Eq.(\ref{1re}) leads to%
\begin{equation}
\sqrt{-t}\left[  A^{(N;q_{1},r_{1})}+A^{(N;q_{1}-1,r_{1}+1)}\right]
-M_{2}A^{(N;q_{1}-1,r_{1})}=0,
\end{equation}
which is the generalization of Eq.(3.90) in \cite{LY} for mass level
$M_{2}^{2}=4$ to arbitrary mass levels $M_{2}^{2}=2(N-1)$. Incidentally, one
should keep in mind that the recurrence relations among RSSA are valid only in
the Regge limit. We give one example to illustrate the calculation. By using
Eq.(\ref{Re1}) and Eq.(\ref{Re2}), we have
\begin{align}
\left(  c-b_{1}-b_{2}\right)  F_{1}\left(  a;b_{1},b_{2};c+1;x,y\right)
-cF_{1}\left(  a;b_{1},b_{2};c;x,y\right)   &  \nonumber\\
+b_{1}F_{1}\left(  a;b_{1}+1,b_{2};c+1;x,y\right)  +b_{2}F_{1}\left(
a;b_{1},b_{2}+1;c+1;x,y\right)   &  =0.
\end{align}
Then with Eq.(\ref{Re3}) and Eq.(\ref{Re4}), we obtain%
\begin{align}
\left(  c-b_{1}-b_{2}\right)  yF_{1}\left(  a;b_{1}-1,b_{2};c;x,y\right)   &
\nonumber\\
+\left[  \left(  a-b_{1}-b_{2}\right)  xy-\left(  c-2b_{1}-b_{2}\right)
y+b_{2}x\right]  F_{1}\left(  a;b_{1},b_{2};c;x,y\right)   &  \nonumber\\
+b_{1}\left(  x-1\right)  yF_{1}\left(  a;b_{1}+1,b_{2};c;x,y\right)
+b_{2}x\left(  y-1\right)  F_{1}\left(  a;b_{1},b_{2}+1;c;x,y\right)   &
=0,\\
\left(  c-b_{1}-b_{2}\right)  xF_{1}\left(  a;b_{1},b_{2}-1;c;x,y\right)   &
\nonumber\\
+\left[  \left(  a-b_{1}-b_{2}\right)  xy-\left(  c-b_{1}-2b_{2}\right)
x+b_{1}y\right]  F_{1}\left(  a;b_{1},b_{2};c;x,y\right)   &  \nonumber\\
+b_{1}\left(  x-1\right)  yF_{1}\left(  a;b_{1}+1,b_{2};c;x,y\right)
+b_{2}x\left(  y-1\right)  F_{1}\left(  a;b_{1},b_{2}+1;c;x,y\right)   &
=0.\label{bb}%
\end{align}
Finally by Combining Eq.(\ref{1re}) and Eq.(\ref{bb}), and taking the leading
term of $s$ in the Regge limit, we end up with the recurrence relation for
$b_{2}$%
\begin{align}
cx^{2}F_{1}\left(  a;b_{1},b_{2};c;x,y\right)   &  \nonumber\\
+\left[  \left(  a-b_{1}-b_{2}-1\right)  xy^{2}+cx^{2}-2cxy\right]
F_{1}\left(  a;b_{1},b_{2}+1;c;x,y\right)   &  \nonumber\\
-\left[  \left(  a+1\right)  x^{2}y-\left(  a-b_{2}-1\right)  xy^{2}%
-cx^{2}+cxy\right]  F_{1}\left(  a;b_{1},b_{2}+2;c;x,y\right)   &  \nonumber\\
-\left(  b_{2}+2\right)  x\left(  x-y\right)  yF_{1}\left(  a;b_{1}%
,b_{2}+3;c;x,y\right)   &  =0,
\end{align}
which leads to a recurrence relation for RSSA at arbitrary mass levels%
\begin{align}
\tilde{t}^{\prime2}A^{(N;q_{1},r_{1})} &  \nonumber\\
+\left[  \tilde{t}^{\prime2}+\tilde{t}\left(  t-2\tilde{t}^{\prime}%
-2q_{1}-2r_{1}+4\right)  \right]  \left(  \frac{\frac{\tilde{t}^{\prime}%
}{2M_{2}}}{\sqrt{-t}}\right)  A^{(N;q_{1},r_{1}+1)} &  \nonumber\\
+\left[  \tilde{t}^{\prime2}-\tilde{t}^{\prime}\left(  \tilde{t}+t\right)
+\tilde{t}\left(  t-2r_{1}+4\right)  \right]  \left(  \frac{\frac{\tilde
{t}^{\prime}}{2M_{2}}}{\sqrt{-t}}\right)  ^{2}A^{(N;q_{1},r_{1}+2)} &
\nonumber\\
-2\left(  r_{1}-2\right)  \left(  \tilde{t}^{\prime}-\tilde{t}\right)  \left(
\frac{\frac{\tilde{t}^{\prime}}{2M_{2}}}{\sqrt{-t}}\right)  ^{3}%
A^{(N;q_{1},r_{1}+3)} &  =0.
\end{align}
More higher recurrence relations which contain general number of $l\geq3$
Appell functions can be found in \cite{Wang}.

\section{Conclusion}

In this paper, we show that open bosonic RSSA can be expressed in terms of one
single Appell function $F_{1}$ in the Regge limit. This result enables us to
derive recurrence relations among RSSA at arbitrary mass levels. In addition,
we show that these recurrence relations of RSSA are so powerful that one can
solve them and all RSSA can be expressed in terms of one single amplitude. All
these results are dual to high energy symmetries of fixed angle string
scattering amplitudes conjectured by Gross in 1988 \cite{Gross} which were
explicitly proved in \cite{ChanLee1,ChanLee2,CHLTY,PRL,susy} previously.

Since it was shown that \cite{sl5c} the Appell function $F_{1}$ are basis
vectors for models of irreducible representations of $sl(5,C)$ algebra, it
seems reasonable to believe that the spacetime symmetry of Regge string theory
is closely related to $SL(5,C)$ non-compact group. In particular, the
recurrence relations of RSSA studied in this paper are related to the
$SL(5,C)$ group as well. Further investigation remains to be done and more
evidences need to be uncovered.

\section{Acknowledgments}

We thank Chung-I Tan for helpful discussion. This work is supported in part by
the National Science Council, 50 billions project of Ministry of Education,
National Center for Theoretical Sciences and S.T. Yau center of NCTU, Taiwan.

\end{document}